\documentclass[final,5p,times,twocolumn,numbers]{elsarticle}
\makeatletter
\AtBeginDocument{%
    \def\NAT@numbers{1}%
    \def\NAT@sort{1}%
    \def\NAT@cmprs{1}%
}
\makeatother

\usepackage{amssymb}
\usepackage{graphicx} % Required for inserting images
\graphicspath{{figs/}}
\usepackage{hyperref}
\hypersetup{hidelinks}
\usepackage{amsmath}
\usepackage{mathtools}

\journal{Physics Letters B}

\begin{document}

\begin{frontmatter}

%% Title, authors and addresses

%% use the tnoteref command within \title for footnotes;
%% use the tnotetext command for theassociated footnote;
%% use the fnref command within \author or \affiliation for footnotes;
%% use the fntext command for theassociated footnote;
%% use the corref command within \author for corresponding author footnotes;
%% use the cortext command for theassociated footnote;
%% use the ead command for the email address,
%% and the form \ead[url] for the home page:
%% \title{Title\tnoteref{label1}}
%% \tnotetext[label1]{}
%% \author{Name\corref{cor1}\fnref{label2}}
%% \ead{email address}
%% \ead[url]{home page}
%% \fntext[label2]{}
%% \cortext[cor1]{}
%% \affiliation{organization={},
%%            addressline={}, 
%%            city={},
%%            postcode={}, 
%%            state={},
%%            country={}}
%% \fntext[label3]{}

\title{Efficient Analysis of Numerous Galactic Compact Binaries via the Free-bin Scheme}

%% use optional labels to link authors explicitly to addresses:
%% \author[label1,label2]{}
%% \affiliation[label1]{organization={},
%%             addressline={},
%%             city={},
%%             postcode={},
%%             state={},
%%             country={}}
%%
%% \affiliation[label2]{organization={},
%%             addressline={},
%%             city={},
%%             postcode={},
%%             state={},
%%             country={}}

\author[a]{Ya-Nan Li}
\author[a]{En-Kun Li}
\author[a]{Yi-Ming Hu\corref{cor1}}
\ead{huyiming@mail.sysu.edu.cn}
\cortext[cor1]{Corresponding author.}
\affiliation[a]{organization={MOE Key Laboratory of TianQin Mission, TianQin Research Center for Gravitational Physics $\&$ School of Physics and Astronomy, Frontiers Science Center for TianQin, Gravitational Wave Research Center of CNSA, Sun Yat-sen University (Zhuhai Campus)},%Department and Organization
            city={Zhuhai},
            postcode={519082}, 
            state={Guangdong},
            country={People's Republic of China}}

\begin{abstract}
%% Text of abstract
Global analysis of Galactic compact binaries (GCBs) for space-based gravitational-wave detectors is computationally demanding in the high-frequency region (above $\sim$10 mHz), where resolvable signals are sparse but frequency samples are numerous. We propose a \textbf{free-bin} scheme that evaluates the nontrivial likelihood only in movable frequency windows centered on sampled signal frequencies. The method preserves the joint-likelihood structure by replacing only the signal-containing regions of a precomputed null likelihood, and avoids the fixed-bin edge effect by construction because the windows follow the sampled frequencies. In a simulated TianQin dataset containing 193 GCBs, 192 injected sources are recovered with waveform fitting factors above 0.99. 
% The recovered posterior samples are consistent with the injected parameters. Extensive numerical comparisons show that the free-bin likelihood closely approximates the conventional full-band likelihood without detectable bias in the tested inference. 
Numerical comparisons demonstrate that the free-bin and conventional full-band likelihood implementations yield identical results when the same finite-length local waveforms and numerical settings are used. 
Computational benchmarks show that the acceleration is primarily controlled by the fraction of the frequency domain covered by the merged signal windows and decreases as this coverage increases.
\end{abstract}

%%Graphical abstract
%\begin{graphicalabstract}
%\includegraphics{grabs}
%\end{graphicalabstract}

%%Research highlights
%\begin{highlights}
%\item Research highlight 1
%\item Research highlight 2
%\end{highlights}

\begin{keyword}
%% keywords here, in the form: keyword \sep keyword, up to a maximum of 6 keywords
gravitational waves \sep Galactic compact binaries \sep joint analysis \sep likelihood acceleration

%% PACS codes here, in the form: \PACS code \sep code

%% MSC codes here, in the form: \MSC code \sep code
%% or \MSC[2008] code \sep code (2000 is the default)

\end{keyword}

\end{frontmatter}

%\tableofcontents

%% \linenumbers

%% main text

\section{Introduction}
\label{sec:introduction}

The data streams of future space-based gravitational-wave detectors, such as TianQin \cite{TianQin:2020hid,2015TianQin}, LISA \cite{Karsten_Danzmann_1996, LISA:2017pwj}, and Taiji \cite{Hu:2017mde, Luo:2021qji}, will contain many overlapping gravitational-wave (GW) signals and instrumental noise \cite{Li_2025, Li:2023szq, Arnaud:2006gm, Arnaud:2007jy, Babak:2008aa, MockLISADataChallengeTaskForce:2009wir, Ren:2023yec, Baghi:2022ucj}. A global analysis is therefore needed to fit detectable sources and noise simultaneously \cite{Cornish:2005qw}. Galactic compact binaries (GCBs) are a central part of this problem: the Milky Way is expected to contain approximately $10^7$ GCBs in the millihertz band, with tens of thousands resolvable and the rest forming a confusion foreground \cite{Cornish:2017vip,Nelemans:2001hp}.

Bayesian Markov chain Monte Carlo (MCMC) methods \cite{Hastings1970}, especially reversible-jump Markov chain Monte Carlo (RJMCMC) \cite{RJMCMC}, have been established as effective tools for trans-dimensional gravitational-wave data analysis \cite{Crowder:2006eu, Littenberg:2011zg, Li:2025eog}. Building on these methods, global analysis pipelines have been developed to infer source catalogs from the data stream \cite{Littenberg:2023xpl,Strub:2024kbe,Katz,Deng:2025wgk}. A major computational bottleneck is likelihood evaluation in the large parameter space. TianQin is expected to detect tens of thousands of resolvable GCBs in [$10^{-4}$ Hz, 0.5 Hz], leading to a parameter-space dimension of $\mathcal{O}(10^5)$. Reducing the cost of likelihood evaluation is therefore important for accelerating global analysis.

Global GCB pipelines typically divide the frequency band into narrow bins and analyze the bins in parallel \cite{Littenberg:2023xpl,Strub:2024kbe,Katz,Deng:2025wgk}. This is efficient in the low-frequency millihertz band, where many signals occupy each frequency interval. Above approximately 10 mHz, however, resolvable GCBs are sparse while the number of frequency samples is large. The [0.01 Hz, 0.5 Hz] band contains about 50 times more frequency samples than the [$10^{-4}$ Hz, 0.01 Hz] band, with only $\mathcal{O}(10^1)\sim\mathcal{O}(10^2)$ GCB signals in this region \cite{Yu:2010fq, Korol:2021pun, Lamberts:2019nyk}. Approximately $98\%$ of the frequency samples lie in the high-frequency region, which contains only about $1\%$ of the resolvable GCB population. Most data points are therefore non-signal, yet they are included in iterative likelihood evaluations, resulting in substantial computational waste. Existing pipelines employ non-uniform frequency bins to ensure that each bin contains $\mathcal{O}(10)$ signals in the estimation step \cite{Littenberg:2023xpl, Katz, Deng:2025wgk}. This approach reduces the number of high-frequency bins, thereby lowering the communication cost between bins. Nevertheless, the computational waste associated with the large number of non-signal data points at high frequencies still remains. Trans-dimensional Bayesian global analysis therefore has potential for further acceleration in this broad but sparse regime.

This study proposes a \textbf{free-bin} scheme to mitigate this computational waste in the high-frequency region. For each trial state in the RJMCMC, the scheme places movable frequency windows around the sampled source frequencies, merges overlapping windows, evaluates the signal likelihood only inside the merged windows, and reuses a precomputed null likelihood elsewhere. The method therefore keeps the joint-likelihood structure of the high-frequency data while substantially reducing the number of data points that enter the nontrivial likelihood evaluation. In addition, it eliminates the edge effects encountered in conventional global analyses.

This work will constitute an important component of TianQin global analysis. Without altering the overarching Bayesian framework or sacrificing the advantages of RJMCMC, it substantially reduces the likelihood-evaluation cost in the sparse high-frequency regime. The speedup depends primarily on the fraction of the frequency domain covered by the active signal windows.

The paper is organized as follows. Sec.~\ref{sec:method} introduces the quasi-monochromatic GCB waveform and the trans-dimensional sampler. Section~\ref{sec:scheme} presents the free-bin likelihood-evaluation scheme for the high-frequency region. Its performance is tested in Sec.~\ref{sec:implement}, and Sec.~\ref{sec:sum} summarizes the results.

\section{Methodology}
\label{sec:method}

\subsection{GCB waveform}\label{sec:waveform}
GCBs are the most numerous sources in the millihertz band of the Milky Way, with tens of thousands expected to be resolvable. These compact binaries, such as double white dwarfs, are nearly circular and can persist for millions of years. Over the TianQin observation time considered here, their GW signals are quasi-monochromatic and can be written in the source frame as \cite{10.1063/1.2405109, PhysRev.131.435}
\begin{equation}
\begin{aligned}
h_{+}(t) &=A_{+} \cos \Phi(t)=h_{0} \frac{1+\cos ^{2} \iota}{2} \cos \Phi(t) \\
h_{\times}(t) & =A_{\times} \sin \Phi(t)=h_{0} \cos \iota \sin \Phi(t) \\
h_{0} & =\frac{4\left(G \mathcal{M}_{c}\right)^{5 / 3}}{c^{4} D_{L}}(\pi f)^{2 / 3} \\
\Phi(t) & =2 \pi f t+\pi \dot{f} t^{2}+\frac{\pi}{3} \ddot{f} t^{3}+\phi_{0}
\end{aligned}
\end{equation}
where $\iota$ is the inclination angle between the orbital angular momentum and the line of sight, $\mathcal{M}_{c}=\left(m_{1} m_{2}\right)^{3 / 5} /\left(m_{1}+m_{2}\right)^{1 / 5}$ is the chirp mass of the binary with component masses $m_1$ and $m_2$, $D_L$ is the luminosity distance, $f$, $\dot f$, and $\ddot f$ are the GW frequency and its first and second time derivatives, $\phi_0$ is the initial phase, and $h_0$ is the intrinsic amplitude. In the implementation section of this paper, the amplitude parameter is denoted by $A\equiv h_0$.

As the detector orbits the Sun within the ecliptic coordinate system, there is a Doppler modulation 
$\Phi_{D}(t)$ on the phase:
\begin{equation}
\begin{aligned}
\Phi(t) & \rightarrow \Phi(t)+\Phi_{D}(t) \\
\Phi_{D}(t) & =2 \pi(f+\dot{f} t) \frac{R}{c} \cos \beta \cos \left(2 \pi f_{m} t-\lambda\right)
\end{aligned}
\end{equation}
where $f_m = 1\,\mathrm{yr}^{-1}$ is the modulation frequency, $\beta$ and $\lambda$ are the latitude and longitude of the source in ecliptic coordinates, and $R=1\,\mathrm{AU}$ is the semi-major axis of the guiding center of the satellite constellation \cite{Li:2023szq}.

Although a single GCB waveform is inexpensive to evaluate, the repeated generation of many waveforms is still costly in a global analysis. We use the fast/slow decomposition method (fastGB) to generate quasi-monochromatic waveforms efficiently \cite{fastgb}. This method decomposes the GCB waveform into a fast carrier term that contains the base frequency $f_0$ and a slow modulation term that contains the remaining parameters. The slow term can be down-sampled while retaining the relevant waveform information, and the fast term can be transformed analytically to the frequency domain. In the free-bin implementation, since the sampled $f_0$ already determines the location of the local computational window, the down-sampled slow-term waveform can be used directly within this window without constructing and positioning the corresponding full-band waveform array. This removes part of the associated array-handling overhead cost.
In the conventional full-band implementation used for comparison, the same downsampled fastGB waveform is positioned within a full-band frequency array, with zero values outside its finite support. Therefore, the two implementations use identical waveform values when the same fastGB sampling settings are adopted.

After the TianQin detector response, the GCB signal in the frequency domain is no longer single-peaked. In addition to the Doppler modulation caused by the detector motion around the Sun, the detector's orbit around the Earth produces a response-induced frequency splitting \cite{Hu:2018yqb}. The TianQin response produces two sidebands separated from $f_0$ by
$f_{\rm resp}=2f_{\rm orb}\simeq6.34\times10^{-6}\ {\rm Hz}$. 
% A width of approximately $4 f_{\rm resp}$ provides a characteristic baseline window for sources below $\sim30$ mHz, which contains the two response peaks and the small additional broadening from Doppler modulation and intrinsic frequency evolution. In the actual implementation, however, the computational window is matched to the width of the locally generated slow-term waveform. Below 30 mHz, this width is approximately $4f_{\rm resp}$. Above 30 mHz, the window width is doubled because stronger frequency evolution at higher frequencies could broaden the signal beyond the $4f_{\rm resp}$ range. This treatment prevents substantial power leakage for high-frequency signals, while retaining narrower windows in the more densely populated region below 30 mHz to reduce the likelihood-evaluation cost.
A width of approximately $4 f_{\rm resp}$ provides a characteristic description of the frequency-domain extent of GCB signals whose intrinsic frequency evolution does not obscure the response-induced splitting. In the actual implementation, however, the computational window is determined by the length of the locally generated slow-term waveform, rather than by an additional manually imposed truncation.

The fastGB implementation adopts an empirical frequency-dependent prescription for the number of downsampled points. For a three-month observation, the waveform contains 256 frequency samples (corresponding to a width of $\sim 4 f_{\rm resp}$) between 10 and 30 mHz, 512 samples ($\sim 8 f_{\rm resp}$) above 30 mHz, and 1024 samples ($\sim 16 f_{\rm resp}$) above 100 mHz. Since the frequency resolution remains unchanged, increasing the number of sampling points also increases the frequency-domain extent of the output waveform.
The 30 mHz and 100 mHz thresholds are empirical settings of fastGB, rather than theoretically derived boundaries of the free-bin scheme. Stronger intrinsic frequency evolution may require additional sampling points to avoid excessive waveform distortion. Therefore, the appropriate sampling length depends not only on the signal frequency but also on its frequency evolution. Users can double the number of sampling points through a parameter called \texttt{oversample}.

\subsection{Trans-dimensional MCMC Sampling}

In global analysis of GCB signals, the number of sources is unknown \emph{a priori} and their waveforms overlap significantly in the data stream. RJMCMC \cite{RJMCMC} addresses this by enabling trans-dimensional sampling: the sampler can propose jumps between models with different numbers of signals and naturally favors those that better fit the data. This simultaneously performs parameter estimation and model selection without fixing the source count beforehand.

In each iteration, a new point $\theta^*$ in parameter space is proposed from the current state $\theta_i$ according to a proposal distribution $q(\theta^* \mid \theta_i)$. The posterior probability is
\begin{equation}
p(\theta \mid D) \propto p(D \mid \theta)\, p(\theta),
\end{equation}
where $p(D \mid \theta)$ is the likelihood and $p(\theta)$ the prior. The Metropolis--Hastings acceptance probability
\begin{equation}
\alpha(\theta_i, \theta^*) = \min\!\left(1, \frac{p(\theta^* \mid D)}{p(\theta_i \mid D)} \frac{q(\theta_i \mid \theta^*)}{q(\theta^* \mid \theta_i)}\right)
\end{equation}
determines whether the proposed state is accepted. The chain converges to the target posterior provided the detailed-balance condition is satisfied.

Because the main goal of this paper is to accelerate likelihood computation, we describe its calculation in detail in Sec.~\ref{sec:likelihood}. In this work, we employ the Python package $\mathtt{Eryn}$ \cite{Eryn, michael_katz_2023_7705496}, which provides an implementation of RJMCMC.

\subsection{Free-bin Scheme}\label{sec:scheme}
\subsubsection{The Likelihood Function}\label{sec:likelihood}

The observed TianQin strain data consist of the superposition of GW signals and instrumental noise, $d=\sum_{i=1}^{N}h_i(\theta)+n$, where $N$ is the number of signals. The goal is to infer a set of waveforms whose residual,
$r(f;\theta)=d(f)-\sum_{i=1}^{N}h_i(f;\theta)$, is consistent with the noise model. Assuming stationary Gaussian noise, the frequency-domain likelihood over a frequency set $\mathcal{F}$ is
\begin{equation}
\begin{aligned}
\mathcal{L}(\theta) 
= \prod_{f\in\mathcal{F}} \frac{1}{\sqrt{(2\pi)^{K} \det\Sigma(f)}} 
\exp \Bigl[ -\frac{1}{2} r^{T}(f;\theta)\Sigma^{-1}(f) r(f;\theta) \Bigr],
\label{equ:likeli}
\end{aligned}
\end{equation}
where $\Sigma(f)$ is the noise covariance matrix for the TDI channels and $K$ is the dimension of the matrix. In this work, we use a fixed analytical noise power spectral density (PSD) \cite{Li:2023szq}, without dynamically updating the noise model.

Instead of using fixed narrow frequency bins and segmentally evaluating the likelihood, we employ several movable ``free'' bins in the high-frequency region. The width of each bin is determined by the bandwidth of a GCB waveform after the detector response \cite{Hu:2018yqb}. 
In each RJMCMC iteration, the bin positions are determined by the current sampled signal frequencies. Within these bins, we evaluate the alternative-hypothesis log-likelihood contribution
\begin{equation}
\begin{aligned}
\ell_1(f;\theta)=C(f)-\frac{1}{2}
&\left[d(f)-\sum_i h_i(f;\theta)\right]^T
\Sigma^{-1}(f) \\
&\times
\left[d(f)-\sum_i h_i(f;\theta)\right],
\end{aligned}
\label{equ:L1}
\end{equation}
where $C(f)=-\frac{1}{2}\log[(2\pi)^K\det\Sigma(f)]$. Outside the bins, the null hypothesis log-likelihood
\begin{equation}
\ell_0(f)=C(f)-\frac{1}{2}d^T(f)\Sigma^{-1}(f)d(f).
\label{equ:L0}
\end{equation}
assumes noise only, is independent of $\theta$ and can be precomputed for all high-frequency samples. 

%%****************************
\begin{figure*}[t]
\centering
\includegraphics[width=1.0\textwidth]{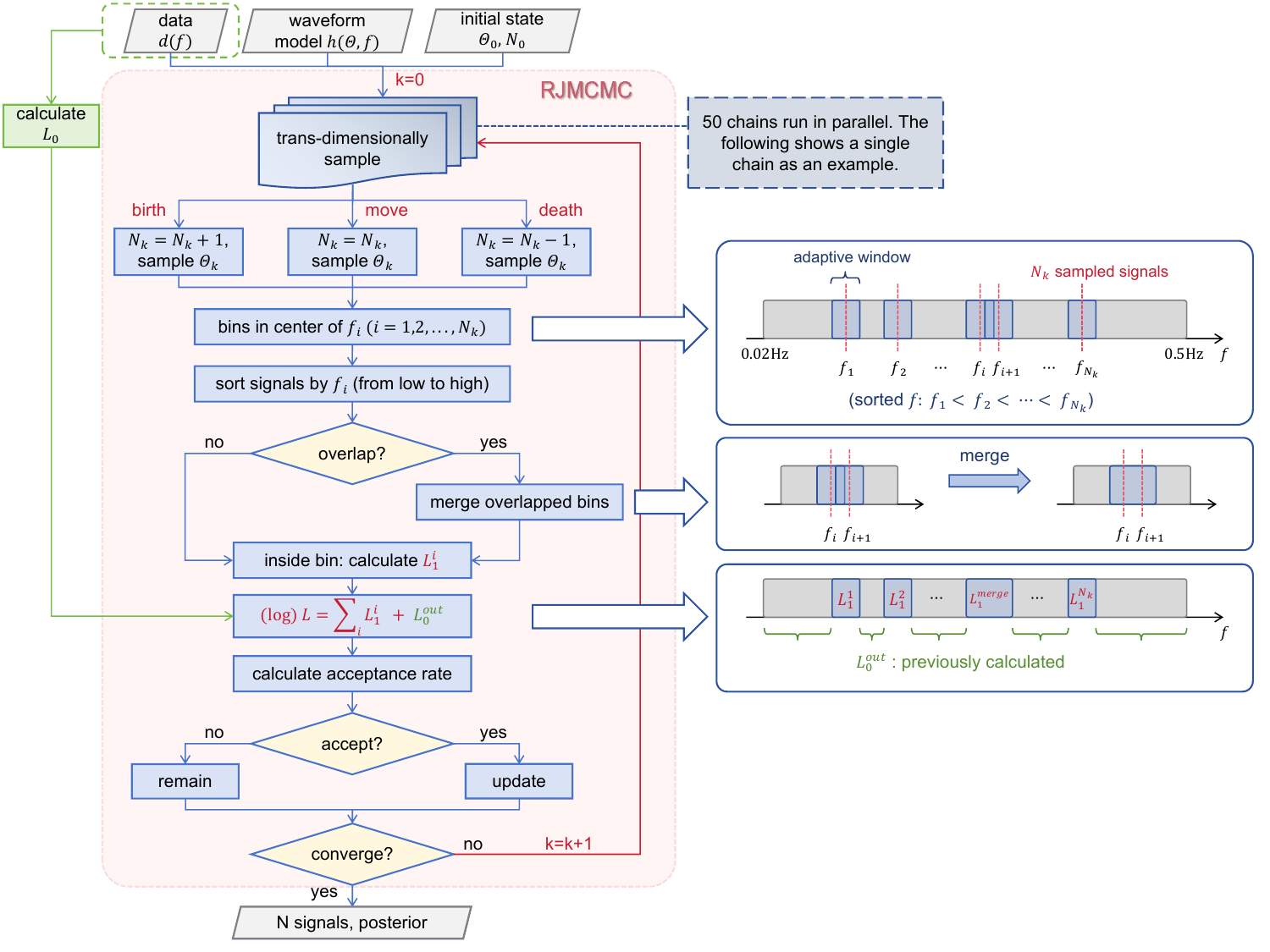}
\caption{Workflow of the RJMCMC sampling and the free-bin likelihood evaluation. Left: $\ell_0(f)$ is precomputed for all high-frequency samples; each RJMCMC state defines movable windows centered on the sampled frequencies; overlapping windows are merged before the likelihood is evaluated. 
% Right: each isolated candidate is assigned an adaptive local window centered on its sampled frequency, with a characteristic baseline width of approximately $4f_{\rm resp}$; overlapping windows are merged and evaluated jointly. 
Right: each isolated candidate is assigned a local computational window centered on its sampled frequency, whose width is determined by the finite-length waveform generated by fastGB; overlapping windows are merged and evaluated jointly. 
The final log-likelihood is obtained by adding the precomputed null contribution and the window corrections in Eq.~(\ref{equ:freebin}).
}
\label{Fig:process}
\end{figure*}
%%************************

Let $\boldsymbol{f}_\theta=\{f_i(\theta)\}_{i=1}^{N(\theta)}$ denote the sampled source frequencies in the current RJMCMC state, and let $B(\boldsymbol{f}_\theta)$ be the union of all movable computational windows centered on these frequencies after overlapping windows have been merged. 
% Under the adopted window choice described in Sec.~\ref{sec:waveform}, signal power outside these windows is approximately negligible. 
For the finite-length fastGB waveform representation described in Sec.~\ref{sec:waveform}, the waveform arrays are identically zero outside the corresponding computational windows.
The free-bin log-likelihood over the high-frequency region $\mathcal{F}_{\rm hi}$ can then be written as
\begin{equation}
\begin{aligned}
\log \mathcal{L}_{\rm hi}^{free}(\theta)
&= \sum_{\nu\in B(\boldsymbol{f}_\theta)} \ell_1(\nu;\theta)
 + \sum_{\nu\in \mathcal{F}_{\rm hi}\setminus B(\boldsymbol{f}_\theta)} \ell_0(\nu) \\
&= \sum_{\nu\in \mathcal{F}_{\rm hi}} \ell_0(\nu)
 + \sum_{\nu\in B(\boldsymbol{f}_\theta)}\left[\ell_1(\nu;\theta)-\ell_0(\nu)\right].
\end{aligned}
\label{equ:freebin}
\end{equation}
Equation~(\ref{equ:freebin}) is the key point of the free-bin scheme: the method does not discard data outside the signal windows, but reuses their precomputed null hypothesis likelihood. 

When multiple sampled signals lie close together in frequency, overlapping bins are merged to avoid redundant computations caused by repeatedly sampling the same signal. Inside a merged window, joint $\ell_1$ is evaluated using the sum of all waveforms whose windows overlap that region. In the notation used in Fig.~\ref{Fig:process}, $L_1^i\equiv\sum_{\nu\in B_i}\ell_1(\nu;\theta)$ denotes the signal-included log-likelihood contribution from the $i$th merged window $B_i$, and $L_0^{\rm out}\equiv\sum_{\nu\in \mathcal{F}_{\rm hi}\setminus B(\boldsymbol{f}_\theta)}\ell_0(\nu)$ denotes the precomputed null contribution from the points outside all windows.

This windowed replacement is well justified. The logarithmic per-frequency contribution to the conventional full-band likelihood Eq.~(\ref{equ:likeli}) is actually the alternative-hypothesis log-likelihood $\ell_1(f;\theta)$, which involves the inner product of the residual $d(f)-\sum_i h_i(f;\theta)$. At frequencies that contain no signal, it is identical to the null-hypothesis log-likelihood $\ell_0(f)$, which involves the inner product of $d(f)$. Since $\sum_i h_i(f;\theta)=0$ at these points,
$
\left\langle d(f)-\sum_i h_i(f;\theta)\middle|d(f)-\sum_i h_i(f;\theta)\right\rangle = 
\left\langle d(f)\mid d(f)\right\rangle$. 
% It should be noted that for the finite windows used here, the remaining outside-window tails are small but not identically zero. Therefore, Eq.~(\ref{equ:freebin}) is a finite-window numerical approximation to the conventional full-band likelihood, rather than an exact equivalence. We quantify the discrepancy between the two log-likelihood values by $\Delta\log\mathcal{L} = |\log\mathcal{L}_{free} - \log\mathcal{L}_{full}|$. Through extensive numerical comparisons, the typical discrepancy is found to be of order $10^{-9}$ in the astrophysically relevant parameter region. Details will be described in Sec.~\ref{sec:valid}. 

It should be noted that the conventional full-band implementation used in our comparisons employs the same finite-length fastGB waveforms as the free-bin implementation. Outside the frequency ranges covered by these waveforms, the corresponding waveform array values are identically zero. Therefore, for the same downsampled fastGB waveforms, the free-bin and full-band likelihoods can be algebraically equivalent.
This equivalence applies to the adopted finite-length waveform representation. If the conventional full-band calculation instead uses a complete frequency-domain waveform without downsampling, the free-bin likelihood is a finite-window numerical approximation.
However, this approximation arises from differences in waveform accuracy caused by different degrees of downsampling, rather than from any bias introduced by the free-bin likelihood itself.

We quantify the numerical discrepancy between the two implementations by $\Delta\log\mathcal{L}=|\log\mathcal{L}_{\rm free}-\log\mathcal{L}_{\rm full}|$. After consistently implementing the floating-point operations, the two likelihood calculations yield identical numerical results for all 20,000 tested individual-source parameter states, as discussed in Sec.~\ref{sec:valid}.

For the three-month dataset used below, the frequency resolution is $\Delta f\simeq1.3\times10^{-7}\ {\rm Hz}$. A baseline window of width $4f_{\rm resp}\simeq2.5\times10^{-5}\ {\rm Hz}$ therefore contains $\sim$200 frequency samples, compared with millions of samples in the full high-frequency region. This substantially reduces the number of data points entering the alternative hypothesis likelihood evaluation.

In fixed-bin analyses, a signal near a bin boundary can have appreciable support in adjacent bins. In the free-bin scheme, the computational window follows the sampled frequency; therefore, an isolated candidate is centered within its own window. The usual fixed-bin edge effect is thus avoided by construction. The merging step also prevents duplicate nearby candidates from gaining artificial likelihood support by being evaluated in separate windows.

\subsubsection{Process}

The implementation follows directly from Eq.~(\ref{equ:freebin}). Before RJMCMC sampling starts, we precompute and store $\ell_0(f)$ for every high-frequency sample. At each RJMCMC step, the current state defines one window centered on each sampled frequency. The windows are sorted by frequency and overlapping windows are merged. In each merged window, $\ell_1(f;\theta)$ is evaluated using the sum of all waveforms assigned to that window; the likelihood update is then obtained by adding $\ell_1-\ell_0$ over the window to the precomputed $\sum_f\ell_0(f)$. The resulting joint log-likelihood is used in the RJMCMC acceptance probability.

\section{Implementation}\label{sec:implement}

\subsection{Validation of Free-bin Likelihood}\label{sec:valid}
% We validate the finite-window approximation by comparing the free-bin (Eq.~(\ref{equ:freebin})) and conventional full-band (Eq.~(\ref{equ:likeli})) likelihoods over a broad set of parameter states. First, we randomly draw 10000 parameter states over the full high-frequency region [10 mHz, 0.5 Hz]. At every state, we independently evaluate the log-likelihood discrepancy $\Delta\log\mathcal L$. The median $\Delta\log\mathcal L$ is of order $10^{-7}$ and the maximum is below $10^{-1}$. The states with $\Delta\log\mathcal{L} > 10^{-4}$ typically have intrinsic amplitudes of order $10^{-20}$ and SNRs of order $10^{4}$. It should be emphasized that such exceptionally strong signals are expected to be very rare in the TianQin data stream.

% Considering a more astrophysically realistic scenario, we adopt all 193 GCBs with $f \ge 10$ mHz from the LISA Data Challenge Sangria (LDC-2a) simulated source catalog \cite{Baghi:2022ucj, le_jeune_2022_7132178}. We generate 10000 parameter states around these sources, constructing a denser population while preserving the main distributional characteristics of the original catalog. These states include posterior-tail and off-source regions of the 193 sources. The median $\Delta\log\mathcal L$ is of order $10^{-9}$ and the maximum is below $10^{-3}$.

We validate the numerical consistency between the free-bin (Eq.~(\ref{equ:freebin})) and conventional full-band (Eq.~(\ref{equ:likeli})) likelihood implementations using two sets of 10,000 individual-source parameter states. Each state contains eight GCB parameters, $(f_0,\dot{f},\beta,\lambda, A,\iota,\psi,\phi_0)$.
The first set consists of 10,000 parameter vectors generated through uniform/log-uniform sampling over broad parameter ranges, including the full high-frequency region [10 mHz, 0.5 Hz]. This set provides broader coverage of the parameter space, including off-source and low-likelihood regions. The second set is generated using kernel density estimation (KDE) based on the parameter distribution of the 193 GCBs with $f\geq10$ mHz from the LISA Data Challenge Sangria (LDC-2a) simulated source catalog \cite{Baghi:2022ucj, le_jeune_2022_7132178}. This KDE-based sampling preserves certain correlations among the source parameters and probes regions associated with the astrophysically motivated source population.

% For both sets of tests, we use simulated TianQin data covering three months of observation, containing GCB signals with the TianQin response and stationary Gaussian instrumental noise in the $A, E, T$ TDI channels. Each individual-source parameter vector is evaluated independently against the same dataset using both likelihood implementations.

We calculate the discrepancy $\Delta\log\mathcal L$ between the two likelihoods at these 20,000 parameter states with the same \texttt{oversample}. The numerical discrepancies were associated with differences in floating-point arithmetic and summation order. To further verify numerical consistency, we unified the frequency ranges, frequency indexing, $A/E$-channel summation order, and other computational details, and adopted the same Neumaier compensated summation algorithm in both implementations, only preserving the core local computation characteristic of the free-bin likelihood and the full-frequency-band computation characteristic of the full-band likelihood. 
With these numerical settings made consistent, the free-bin and full-band likelihood implementations produced identical log-likelihood values for all 20,000 tested individual-source parameter states, with $\Delta\log\mathcal{L}=0$ in every case. This result confirms their numerical equivalence for the same finite-length fastGB waveform representation.
For the overlapping configurations in the 193 sources, the likelihood discrepancies of the joint overlapping states are also zero.

% In addition to the pointwise validation of the likelihood values, we also performed two sampler-level comparisons to demonstrate that the free-bin likelihood does not introduce additional bias into the sampling results. Considering a manageable end-to-end runtime, we performed a test on a computationally manageable $\sim$20 mHz sub-band containing 4 GCB signals \cite{Korol:2017qcx}. We performed the RJMCMC sampling with the Python package $\mathtt{Eryn}$ \cite{Eryn, michael_katz_2023_7705496}, and evolved the free-bin and full-band implementations from the same initial state using identical random-number sequences. Their RJMCMC trajectories remain identical over the tested evolution. After $5\times10^5$ steps, the two implementations yield identical posterior distributions, with a Kullback–Leibler (KL) divergence of 0, indicating no detectable difference between the posterior distributions obtained with the two likelihood implementations.

\subsection{Astrophysically Motivated Case}\label{sec:193s}

For all 193 GCBs with $f \ge 10$ mHz in the LDC-2a catalog \cite{Baghi:2022ucj, le_jeune_2022_7132178}, we generate the signal waveforms including the TianQin response, and inject them into a data stream with stationary Gaussian instrumental noise. This simulated dataset reproduces the expected trend of increasingly sparse signals toward higher frequencies. To suppress the overwhelming laser phase noise, space-based detectors rely on time-delay interferometry (TDI) \cite{Prince:2002hp, Tinto:2020fcc}. We use the idealized orthogonal $A$, $E$, and $T$ TDI channels, for which the instrumental noise is statistically uncorrelated between channels. The off-diagonal elements of the noise covariance matrix $\Sigma$ therefore vanish, which simplifies the likelihood evaluation. The observation time is three months, and the analytical PSD is fixed during sampling rather than fitted dynamically.

In this implementation, we set the fastGB downsampling parameter \texttt{oversample}=1, retaining the default number of downsampled points internally determined by fastGB, as described in Sec.~\ref{sec:waveform}. Below 30 mHz, the waveform width is $256\Delta f$ ($\sim 4f_{\rm resp}$), while above 30 mHz, the waveform width is $512\Delta f$ ($\sim 8f_{\rm resp}$). For the 193 sources in this example, these settings ensure that the retained power fraction exceeds 99\% for every signal.

We perform the RJMCMC sampling with $\mathtt{Eryn}$ \cite{Eryn, michael_katz_2023_7705496}. The proposal distribution combines an in-model Gaussian move, which updates the eight parameters of existing signals, with an RJ move that proposes additions or removals of signals and changes the model dimension. We run 50 walkers for $10^6$ steps, corresponding to $5\times10^7$ walker updates before thinning, and retain 100,000 thinned posterior samples.

Because the raw RJMCMC chains contain samples of varying dimensionality, post-processing is required to produce a definite source catalog with individual posterior distributions for each signal. We follow the catalog-production method of Littenberg et al. \cite{Littenberg:2020bxy}, grouping signals into entries by calculating their waveform fitting factor:
\begin{equation}
FF \equiv \frac{\langle\boldsymbol{h} \mid \boldsymbol{g}\rangle}{\sqrt{\langle\boldsymbol{h} \mid \boldsymbol{h}\rangle\langle\boldsymbol{g} \mid \boldsymbol{g}\rangle}}.
\end{equation}
$\boldsymbol h$ and $\boldsymbol g$ are two different signal waveforms, and the inner product is
\begin{equation}
\langle \boldsymbol{h} \mid \boldsymbol{g} \rangle
=
4\,\mathrm{Re}
\int_{f_{\min}}^{f_{\max}}
\frac{\tilde h(f)\tilde g^{*}(f)}
     {S_n(f)}
\,df.
\end{equation}
${S_n(f)}$ is the one-sided noise power spectral density (PSD) of the detector. The tilde indicates the Fourier transform of the corresponding time-domain signal and the superscript asterisk denotes complex conjugation.
We use $FF>0.9$ as the matching threshold for identifying a recovered entry with an injected source. A pre-filter that keeps only pairs within one signal width in frequency is used only to accelerate catalog construction. Besides, the first 50\% of the samples are discarded and the remaining samples are further thinned by 10. The catalog-production time is reduced to approximately 30 minutes.

\begin{figure*}[!ht]
\centering
\includegraphics[width=1.0\textwidth]{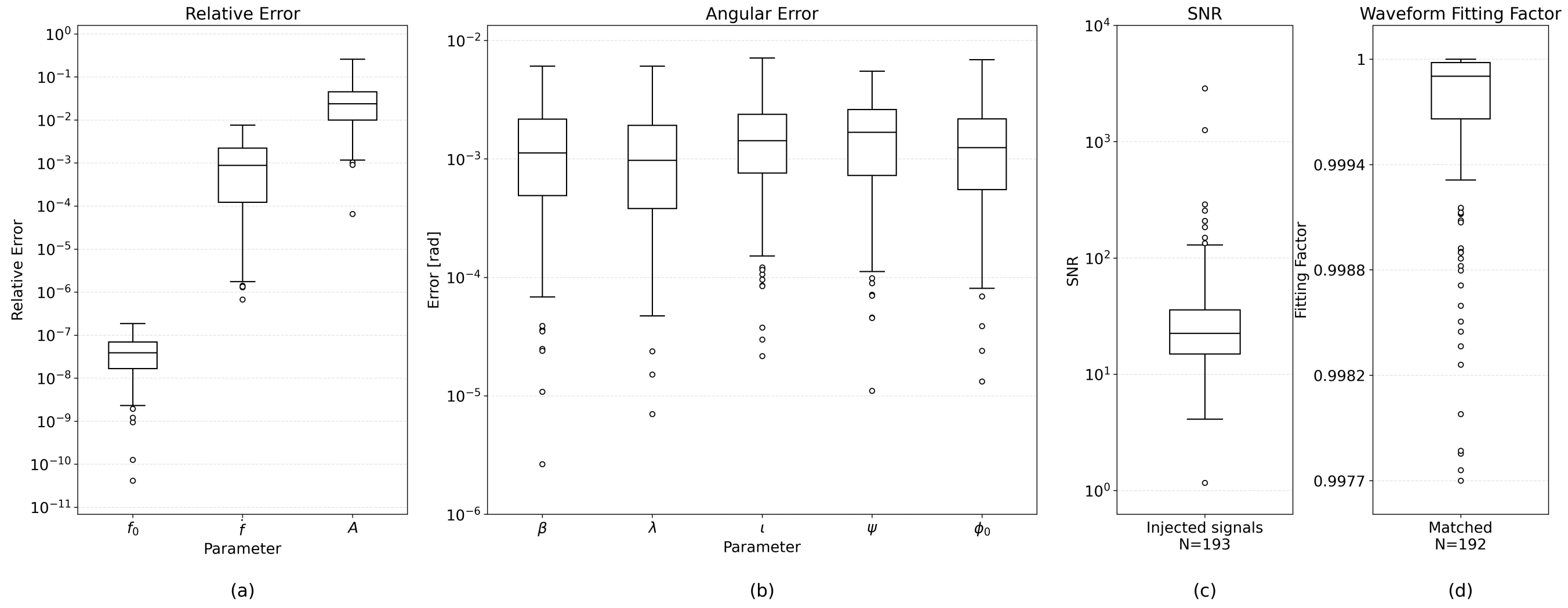}
\caption{Box plots of the parameter-estimation results. The two subplots on the left show the parameter errors for the 192 matched entries. Subplot (a) presents the absolute relative errors $|(\theta_{\rm rec}-\theta_{\rm inj})/\theta_{\rm inj}|$ for the non-angular parameters ($f_0, \dot{f}, A$). Subplot (b) displays the wrapped absolute angular errors $|\mathrm{wrap}(\theta_{\rm rec}-\theta_{\rm inj})|$ in radians for the angular parameters ($\beta, \lambda, \iota, \psi, \phi_0$), where $\psi$ is the polarization angle. Subplot (c) illustrates the SNR distribution of the 193 injected signals, with the lowest value being 1.16. Subplot (d) presents the fitting factors for the 192 matched signals, all of which are greater than 0.99.}
\label{Fig:box}
\end{figure*}

After classification, we obtain 193 catalog entries. Taking the median parameters of each entry as the recovered values, 192 entries match injected sources; in practice, all matched entries have $FF>0.99$. The fitting-factor distribution is shown in Fig.~\ref{Fig:box}(d). One injected source is missed; it has the lowest signal-to-noise ratio, ${\rm SNR}=1.16$, and is the weakest source in the catalog, as indicated by the lowest point in Fig.~\ref{Fig:box}(c). The remaining unmatched catalog entry is a low-occupancy spurious entry near the highest-SNR source.

\begin{figure*}[!ht]
\centering
\includegraphics[width=1.0\textwidth]{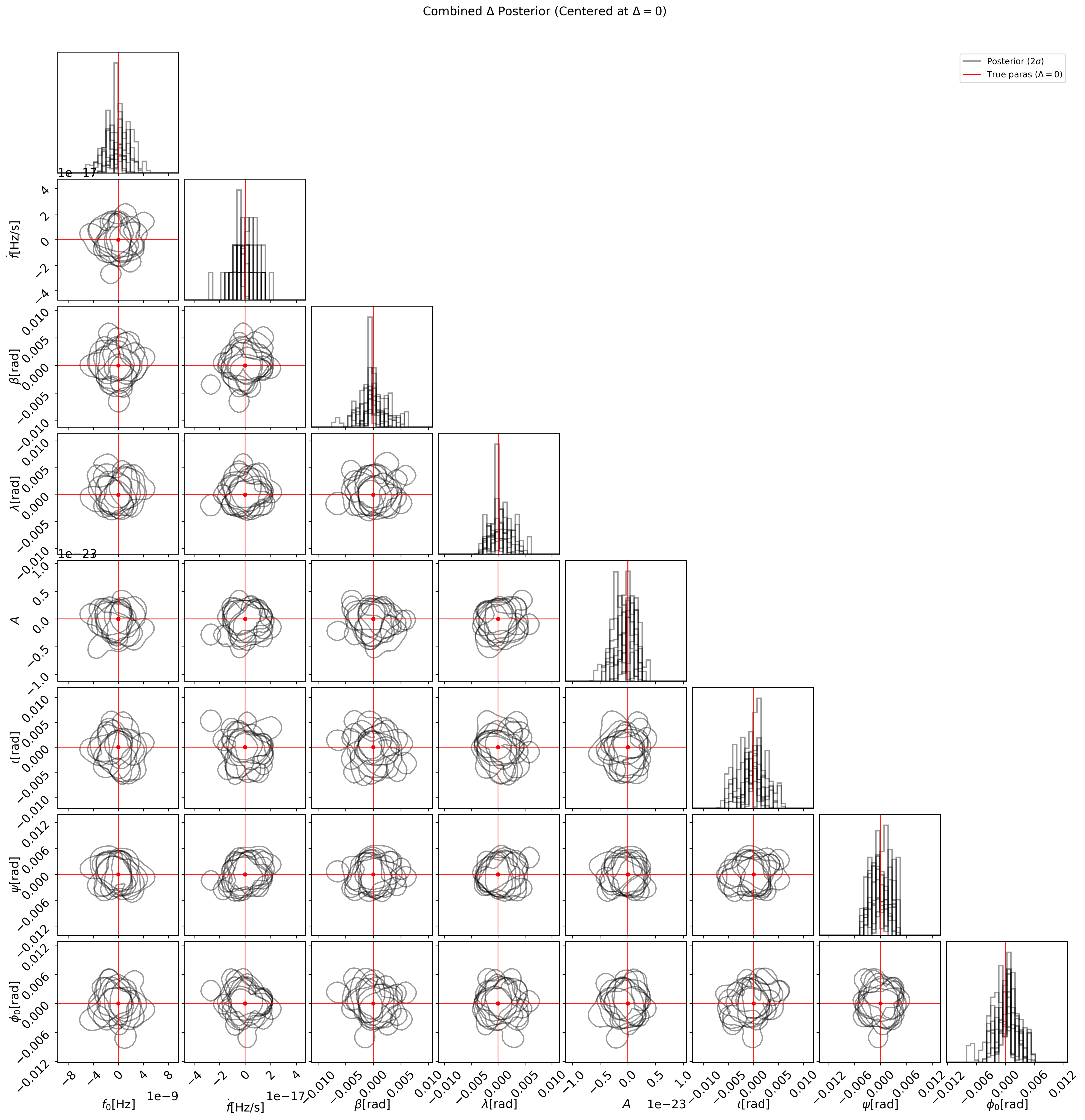}
\caption{Posterior distributions of 10 randomly selected recovered signals. The plotted quantities are recovered-minus-injected parameter errors, so the injected value is zero for every parameter. The black contours indicate the $2\sigma$ posterior credible regions for the plotted marginalized distributions.}
\label{Fig:join}
\end{figure*}

Figure \ref{Fig:join} shows the posterior distributions of 10 randomly selected recovered signals. Because the signals occupy different locations in parameter space, we plot recovered-minus-injected parameter errors so that the injected value is zero for every parameter. The black contours indicate the $2\sigma$ posterior credible regions for the plotted marginalized distributions.

To quantitatively demonstrate the parameter estimation results, we plot box plots of the parameter errors for the 192 matched entries, as shown in Figure \ref{Fig:box}(a) and (b). The median errors are below $10^{-2}$ for all parameters except the amplitude, whose median relative error is $2.39\times10^{-2}$; the median relative error in $f_0$ is $3.91\times10^{-8}$.

\subsection{Acceleration Performance}\label{sec:speedup}

\begin{figure}[!h]
\centering
\includegraphics[width=0.45\textwidth]{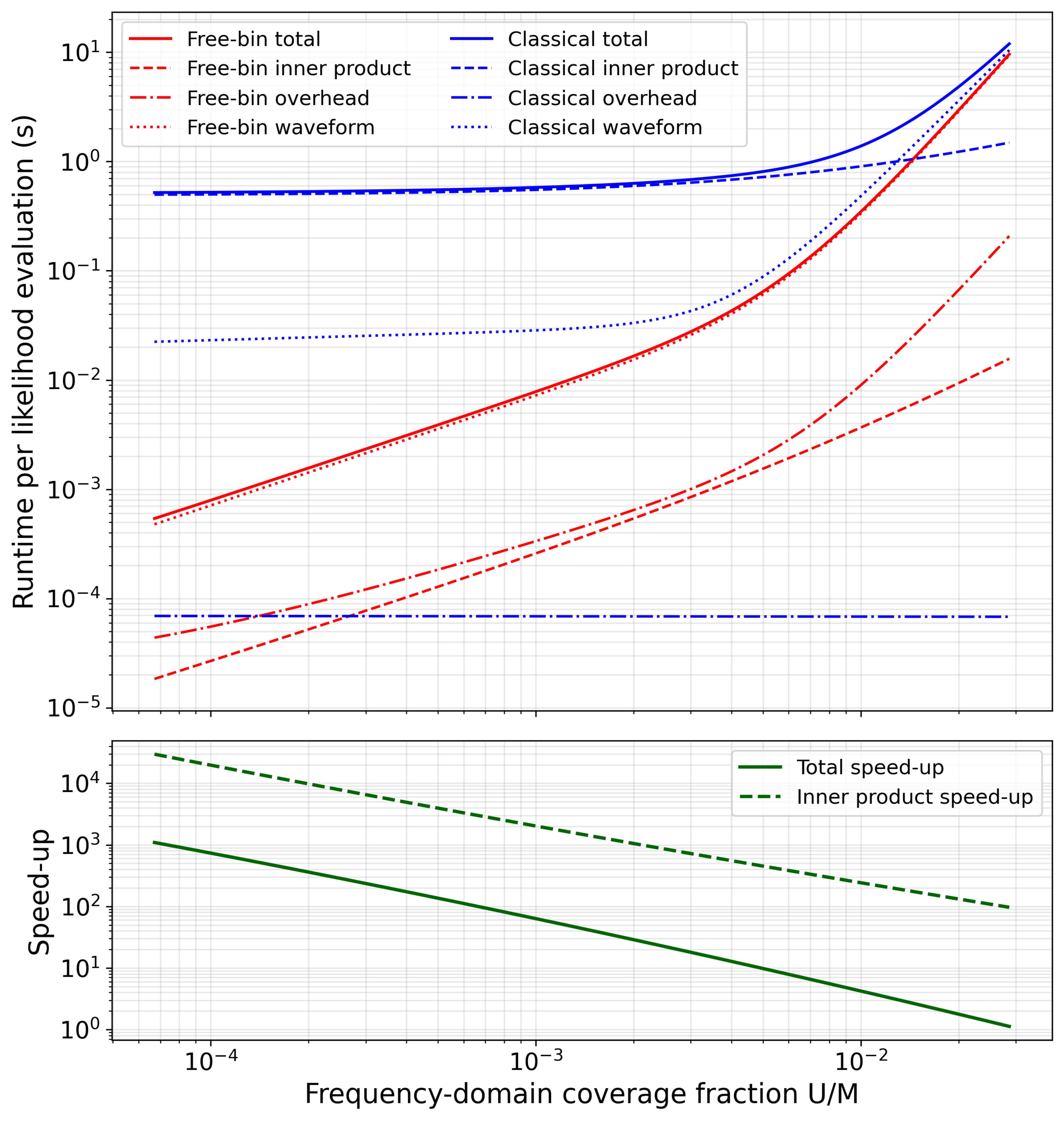}
\caption{Computational scaling of the free-bin likelihood. Upper: Runtime per likelihood evaluation for the waveform-generation, inner-product, overhead, and total components of the free-bin and conventional full-band implementations as a function of the frequency-domain coverage fraction $U/M$. Lower: Corresponding speedup ratios of the inner-product component and the total likelihood evaluation. The curves are quadratic polynomial fits to the measured timing points and are shown to illustrate the trends.}
\label{Fig:timing}
\end{figure}

We evaluate the computational performance of the free-bin scheme against the standard full-band likelihood calculation on the same dataset and hardware. 
% The free-bin scheme accelerates a single likelihood evaluation by a factor of $\sim 10^4$, consistent with the reduced number of frequency samples entering the $\ell_1$ calculation. This factor applies only to likelihood evaluation on a single signal. 
The free-bin scheme achieves a maximum speedup of approximately $10^4$ for the inner-product component of likelihood evaluation in the single-source limit, consistent with the reduced number of frequency samples entering the $\ell_1$ calculation. This acceleration should be distinguished from the speedup of the complete likelihood evaluation, which also includes waveform generation and auxiliary operations. As shown in Fig.~\ref{Fig:timing}, the maximum speedup of the complete likelihood evaluation is of order $10^3$ in the low-coverage regime. 

The end-to-end RJMCMC runtime also includes sampler overhead and bookkeeping. 
In the 193-source case, the entire RJMCMC run finishes in less than two days on two 28-core Intel Xeon Gold 6330 CPUs. 
To estimate the end-to-end acceleration, we also performed an RJMCMC run using the conventional full-band likelihood under the same hardware configuration. Since a complete full-band run is computationally expensive, we compared the wall-clock runtimes of approximately 10,000 sampling steps for each implementation. The free-bin run was $\sim$10 times faster over this interval.
Because the computational cost per RJMCMC step remains approximately stable during the tested evolution, this comparison provides an estimate of the end-to-end speedup for longer runs.

The acceleration achieved by the free-bin scheme depends on the number of signals and the degree of overlap among them.
To quantify the computational scaling, we define the frequency-domain coverage fraction $U/M$, where $M$ is the total number of frequency samples in the high-frequency region and $U$ is the number of unique samples covered by the merged free-bin windows. Unlike the source number $N$, $U/M$ directly measures the fraction of the data entering the nontrivial likelihood calculation. We decompose the cost of one likelihood evaluation as $T_{\rm total}=T_{\rm waveform}+T_{\rm inner}+T_{\rm overhead}$, where the three terms denote waveform generation, likelihood inner-product evaluation, and auxiliary operations. We benchmark these components for source populations ranging from $N=1$ to $10^4$.

Figure~\ref{Fig:timing} shows that the conventional full-band likelihood is dominated by the inner-product calculation, whereas the free-bin scheme strongly suppresses this cost and shifts the dominant contribution to waveform generation. As $U/M$ increases, the free-bin cost approaches the full-band value and both the inner-product and total speedups decrease. The total speedup is lower because waveform generation and auxiliary operations remain. The auxiliary cost in the free-bin likelihood is somewhat larger because of the additional sorting and merging operations. In our 193-source case, $U/M\approx 6\times10^{-3}$, the end-to-end speedup is $\sim$10 times.

A complementary test with $N=100$ fixed while varying the overlap fraction confirms this interpretation: stronger overlap reduces the unique coverage $U$ and therefore increases the inner-product speedup, whereas the total speedup changes only weakly because waveform generation dominates the free-bin runtime.

\section{Summary}\label{sec:sum}
We have presented a \textbf{free-bin} likelihood-evaluation scheme for the joint analysis of sparse high-frequency GCBs in space-based GW detectors. The scheme centers computational windows on the sampled source frequencies, merges overlapping windows, evaluates the signal likelihood only inside these windows, and reuses a precomputed null likelihood elsewhere. This preserves the joint-likelihood structure of the high-frequency region while mitigating the computational waste associated with evaluating signal likelihoods over frequency regions without candidate signals. 
% The resulting free-bin likelihood is a finite-window numerical approximation to the conventional full-band likelihood. Extensive tests over posterior, off-source, and overlapping-source configurations show a typical absolute log-likelihood discrepancy of order $10^{-9}$ in the astrophysically relevant parameter region.

For the same finite-length fastGB waveform representation, the free-bin and conventional full-band likelihoods can be algebraically equivalent. After unifying the float-point numerical implementation details, the two implementations yield identical log-likelihood values for all 20,000 tested individual-source parameter states. The free-bin scheme would be a finite-window numerical approximation if the reference full-band likelihood instead employed a complete frequency-domain waveform without downsampling.

% Controlled full-band/free-bin sampling tests show consistent RJMCMC evolution and posterior inference in the tested subsets, demonstrating that the free-bin likelihood does not introduce additional bias into the sampling results.

In a simulated TianQin dataset with 193 injected GCBs above 10 mHz, the catalog contains 193 entries: 192 entries match injected sources with $FF>0.99$, the weakest injected source is missed with ${\rm SNR}=1.16$, and one unmatched entry is a low-occupancy spurious entry near the highest-SNR source. The posterior distributions are centered close to the injected values, and the recovered parameter errors are small.

Computational benchmarks show that the acceleration is controlled primarily by the merged-window coverage fraction $U/M$. The maximum inner-product speedup reaches approximately $10^4$ in the single-signal limit, whereas the complete likelihood-evaluation speedup reaches approximately $10^3$ in the low-coverage regime. Both decrease as the frequency coverage increases. 
In the astrophysically motivated 193-source implementation, the run with 50 walkers for $10^6$ steps completes in less than two days on two 28-core Intel Xeon Gold 6330 CPUs. The end-to-end time is $\sim$10 times faster than the conventional full-band likelihood implementation under the same hardware configuration.

This work addresses the computational waste in likelihood evaluation arising from the fact that, at high frequencies, $98\%$ of the data points contain only $1\%$ of the signals. It is also, in principle, applicable to the low-frequency region. However, severe source confusion at low frequencies leaves few non-signal data points, resulting in no significant acceleration. The low-frequency region will be handled by another module in the global analysis pipeline.

We also note that 10 mHz is a commonly adopted boundary between the low- and high-frequency regions in current global analysis studies. This is not a strict division, and some researchers instead define frequencies above 18 mHz as the high-frequency region \cite{Deng:2025wgk}. The choice of boundary does not actually affect the implementation of the free-bin scheme. Based on the distribution of GCB signals in the frequency domain, adopting this boundary to separate the low- and high-frequency analyses is reasonable.

The present validation uses idealized stationary Gaussian noise and a fixed analytical PSD. Future work will integrate the scheme into global analysis pipelines and test its performance on realistic TianQin data challenges. By reducing the high-frequency likelihood cost while retaining efficient source identification, the free-bin scheme is expected to become a useful component of the TianQin global data-analysis framework.

\section*{Acknowledgements}

This work has been supported in part by 
the National Key Research and Development Program of China 
(No. 2023YFC2206701) and the Fundamental Research Funds for the
Central Universities, Sun Yat-sen University.
YMH is also supported by the science research grants from the China Manned Space Project.

\section*{Declaration of generative AI and AI-assisted technologies in the manuscript preparation process}

During the preparation of this work the authors used ChatGPT in order to optimize the algorithm code and implement auxiliary functions such as plotting in Python. After using this tool/service, the authors reviewed and edited the content as needed and take full responsibility for the content of the published article.

%% The Appendices part is started with the command \appendix;
%% appendix sections are then done as normal sections
%%\appendix

%%\section{Appendix title 1}
%% \label{}

%%\section{Appendix title 2}
%% \label{}

%% If you have bibdatabase file and want bibtex to generate the
%% bibitems, please use
%%
%%\bibliographystyle{elsarticle-harv} 
\bibliographystyle{elsarticle-num}
\bibliography{refs}

%% else use the following coding to input the bibitems directly in the
%% TeX file.

%%\begin{thebibliography}{00}

%% \bibitem[Author(year)]{label}
%% For example:

%% \bibitem[Aladro et al.(2015)]{Aladro15} Aladro, R., Martín, S., Riquelme, D., et al. 2015, \aas, 579, A101

%%\end{thebibliography}

\end{document}